\documentclass[twocolumn,preprintnumbers,amsmath,amssymb,superscriptaddress]{revtex4}

\usepackage{xcolor}
\usepackage{blindtext}
\usepackage{graphicx}
\usepackage{dcolumn}
\usepackage{bm}
\usepackage{gensymb}
\usepackage{animate}
\usepackage{soul}

\begin{document}


\title{Probing topological protected transport in finite-sized Su-Schrieffer-Heeger chains}

\author{Yu-Han Chang}
\affiliation{Department of Physics, National Chung Hsing University, Taichung 402, Taiwan}

\author{Nadia Daniela Rivera Torres}
\affiliation{Department of Physics, National Tsing Hua University, Hsinchu 300, Taiwan}

\author{Santiago Figueroa Manrique}
\affiliation{Department of Physics, National Tsing Hua University, Hsinchu 300, Taiwan}

\author{Raul A. Robles Robles}
\affiliation{Institute of Photonics Technologies, National Tsing Hua University, Hsinchu 300, Taiwan}

\author{Vanna Chrismas Silalahi}
\affiliation{Department of Physics, National Chung Hsing University, Taichung 402, Taiwan}

\author{Cen-Shawn Wu}
\affiliation{Department of Physics, National Chang-Hua University of Education, Changhua 500, Taiwan}

\author{Gang Wang}
\affiliation{School of Physical Science and Technology, Soochow University, Suzhou 215006, China}

\author{Giulia Marcucci}
\affiliation{Department of Physics, University Sapienza, Rome 00185, Italy}
\affiliation{Institute for Complex Systems, National Research Council (ISC-CNR), Rome 00185, Italy}
\affiliation{Apoha Ltd, 242 Acklam Rd, London W10 5JJ, United Kingdom}

\author{Laura Pilozzi}
\affiliation{Institute for Complex Systems, National Research Council (ISC-CNR), Rome 00185, Italy}
 \affiliation{Research Center Enrico Fermi, Via Panisperna 89a, 00184 Rome, Italy}

\author{Claudio Conti}
\affiliation{Department of Physics, University Sapienza, Rome 00185, Italy}
\affiliation{Institute for Complex Systems, National Research Council (ISC-CNR), Rome 00185, Italy}
\affiliation{Research Center Enrico Fermi, Via Panisperna 89a, 00184 Rome, Italy}

\author{Ray-Kuang Lee}
\email{rklee@ee.nthu.edu.tw}
\affiliation{Department of Physics, National Tsing Hua University, Hsinchu 300, Taiwan}
\affiliation{Institute of Photonics Technologies, National Tsing Hua University, Hsinchu 300, Taiwan}
\affiliation{Physics Division, National Center for Theoretical Sciences, Taipei 10617, Taiwan}
\affiliation{Center for Quantum Technology, Hsinchu 30013, Taiwan}

\author{Watson Kuo}
\email{wkuo@phys.nchu.edu.tw}
\affiliation{Department of Physics, National Chung Hsing University, Taichung 402, Taiwan}
\affiliation{Center for Quantum Technology, Hsinchu 30013, Taiwan}

\begin{abstract}
In order to transport information with topological protection, we reveal and demonstrate experimentally the existence of a characteristic length $L_c$, coined as the {\it transport length}, in the bulk size for edge states in  one-dimensional Su-Schrieffer-Heeger (SSH) chains.
In spite of the corresponding wavefunction amplitude decays exponentially, characterized by the penetration depth $\xi$, the transport between two  edge states  remains possible even when the lattice size $L$ is much larger than the penetration depth, i.e., $\xi \ll L \le L_c$.
Due to the non-zero coupling energy in a finite-size system,  the supported SSH edge states are not completely isolated at the two ends, giving an abrupt change in the wave localization, manifested through the inverse participation ratio to the lattice size.
To verify such a non-exponential scaling factor to the system size, we implement a chain of split-ring resonators and their complementary ones with controllable hopping strengths.
By performing the measurements on the group velocity from the transmission spectroscopy of non-trivially topological edge states with pulse excitations, the transport velocity between two edge states is directly observed with the number of lattices up to $20$.
Along the route to harness topology to protect optical information, our  experimental demonstrations provide a crucial guideline for utilizing photonic topological devices. 
\end{abstract}
\maketitle

\noindent {\it Introduction.}\textemdash
Composited by dimers with staggered hopping amplitudes,  a topological phase transition can be revealed in 
the Su-Schrieffer-Heeger (SSH) model owing to the existence of  Zak phase associated with zero Berry curvature
~\cite{SSH-1, SSH-2}.
By utilizing the SSH model, people have illustrated the difference between bulk and boundary, chiral symmetry, adiabatic equivalence, topological invariants, and bulk–boundary correspondence~\cite{book-SSH}.
Through the analogy in single-particle Hamiltonian, topologically non-trivial zero or $\pi$ modes can also be observed in photonic systems, through a periodically setting on the confining potentials~\cite{OL-Chen, TP-mat, TP-photon, TP-2D, Tao}.
With the topologically protected edge states, we can implement new types of lasing modes~\cite{laser1,laser} and optical control~\cite{control, nlin-SSH} even under continuous deformations.


As  the edge modes in a one-dimensional (1D) SSH model are zero-dimensional, by definition, they do not have a group velocity. 
However, for the practical implementation, instead of infinite chains, only a finite number of dimers can be fabricated. 
Moreover, in order to harness topology to protect optical information~\cite{tp-emitter}, a natural question arises on the corresponding  transport properties in the supported non-trivially topological states.
It is the common belief that  the wavefunctions of supported  edge states in a finite-size chain should remain staying localized  strongly  at their respective boundaries and decay exponentially in the bulk, with a penetration depth $\xi$ depending on the contrast between coupling strengths.
On the contrary, in this {\it Letter}, we reveal the existence of a transport length $L_c$ in finite-size SSH chains, which  is much larger than the penetration depth, i.e., $L_c \gg \xi$.
As long as the system size is smaller than this transport length, one can transport information  between two supported edge states with non-trivially topological protection.

By calculating the corresponding  inverse participation ratio (IPR), an abrupt change in the wave (de)localization happens at this critical length scale, with the comparison to the exponential scaling factor to the system size from the topologically trivial states.
Theoretically, the dependence on the lattice size and the contrast between coupling strengths is also derived for the transport length, see Eq. (3) below.
Experimentally,  the existence of non-trivially optical topological edge states in a chain of dimers with split-ring resonator (SRR) and its counterpart, the complementary split-ring resonator (CSRR), is illustrated with  a proper setting on the intra- and inter-cell coupling strengths between SRRs and CSRRs~\cite{Falcone04}.
By extracting the amplitude and phase from the transmission  spectroscopy, a photonic band gap in the passband is measured  when the inversion symmetry is broken~\cite{Shelby01, Linden04}.
The corresponding group velocity of edge states is directly measured with continuous wave and pulse excitations, verifying the transport of edge states when the system size is much larger than the penetration depth.
Our results indicate that  within this transport length scale the supported edge states possess not only  non-trivial topologies, but also a non-exponentially scaling factor in a finite-size SSH model.\\

\begin{figure}[t]
\includegraphics[width=8.4cm]{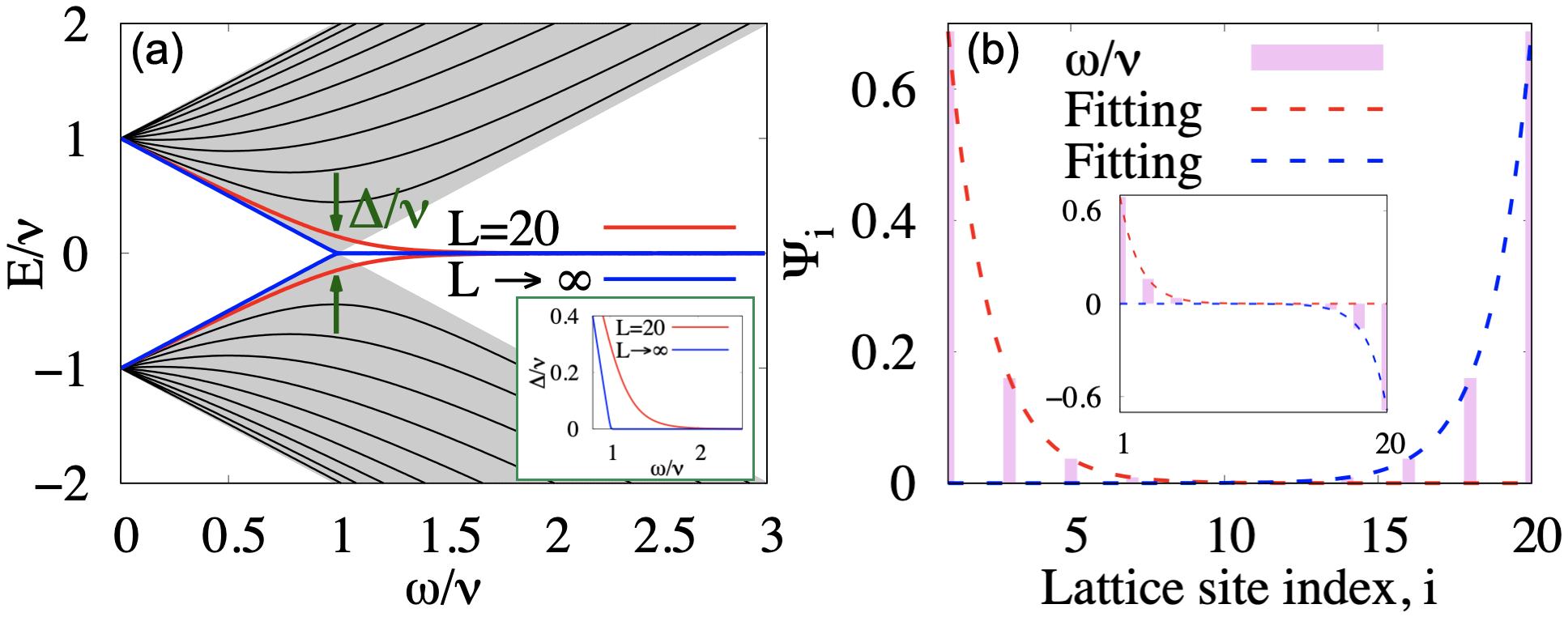}
\caption{(a) Energy bands of edge states for finite-size ($L = 20$) and infinite-size ($L = \infty$) SSH models. The coupling energy $\Delta$ closes ($\Delta = 0$) when $w/v \ge 1$ only for an infinite-size system. However,  for finite-size systems a non-zero value of the energy gap $\Delta$ occurs even when $w/v > 1$, as depicted in the inset. The corresponding edge modes $\Psi_i$ supported in a finite-size SSH model ($L = 20$) are shown in (b), with the amplitude decaying exponentially, i.e., the fitting dashed-curves of $\text{exp}[-\text{i}/\xi]$ and $\text{exp}[-(L-\text{i})/\xi]$ with the lattice site index,  $\text{i}$, and the penetration depth $\xi$.
We want to remark that, instead of the wavefunction pinned to one end, the supported edge state can be symmetric or anti-symmetric combination of the isolated edge modes, as shown in (b) and the inset.
}
\end{figure}

\noindent {\it Finite-size SSH chains.}\textemdash
We consider the SSH model, a  1D tight-binding model with alternating coupling strengths in dimer lattices. 
In the unit cell of ``diatomic basis'', the  staggered coupling strengths, denoted as $v$ and $w$, account for the  intra- and inter-cells, which  lead to a parity effect in the chain length and a gap for the infinite chain case. 
The SSH  model is known for its topological features and shows a transition from a trivial phase to a topological phase as it is tuned crossing $w/v  =1$.

When $w/v \ge1$, the normalized energy of edge modes lies at the band center as shown in Fig. 1(a).
Outside the band center, one recovers typical Bloch solutions as expected for periodic systems.
However, as one can see in Fig. 1(a), the energy gap $\Delta$ closes ($\Delta = 0$) when $w/v \ge 1$ only for an infinite-size system. 
For a finite-size system, such as $L=20$ in our experiments, a non-zero value of the energy difference $\Delta$ occurs for ``near-zero" modes even when $w/v > 1$, as depicted in the inset of Fig. 1(a). 
Here, $L$ denotes the number of lattice sites. Given $w/v>1$, $\Delta$ can be approximated by 
$\Delta=2w (\frac{w}{v})^{-N}[1-(\frac{w}{v})^{-2}]/{[1-(\frac{w}{v})^{-2N}]}$ with $N=L/2$,  the number of unitary cells~\cite{Efremidis}.
Associated with a finite $\Delta$, the corresponding modes are no longer isolated at one end, but become a symmetric or anti-symmetric combination of the isolated edge modes, as depicted in Fig. 1(b). In this case, $\Delta/2$ can be viewed as a coupling energy of the localized modes on each edge, arising from inter-cell coupling $w$ but scaled exponentially by sample size $L$. 

Similar to the edge states in an infinite-size system, the wavefunction amplitude of supported edge modes in a finite-size SSH model decays exponentially, characterized  by the penetration depth $\xi = 2/\ln(w/v)$~\cite{Efremidis}.
As shown in Fig. 1(b), the exponential decaying factors, i.e., $\text{exp}[-i/\xi]$ and $\text{exp}[-(L-{i})/\xi]$ with the site label $i$, give agreement to the wavefunction amplitudes at both ends.
Nevertheless, we want to remark that the penetration length $\xi$, independent of the system size $L$, is only a few lattice size. 
In the following, our experiments work at $w/v = 4.3$, which gives a penetration length $2\, \xi \approx 2.76$ sites.\\

\begin{figure}
\includegraphics[width=8.4cm]{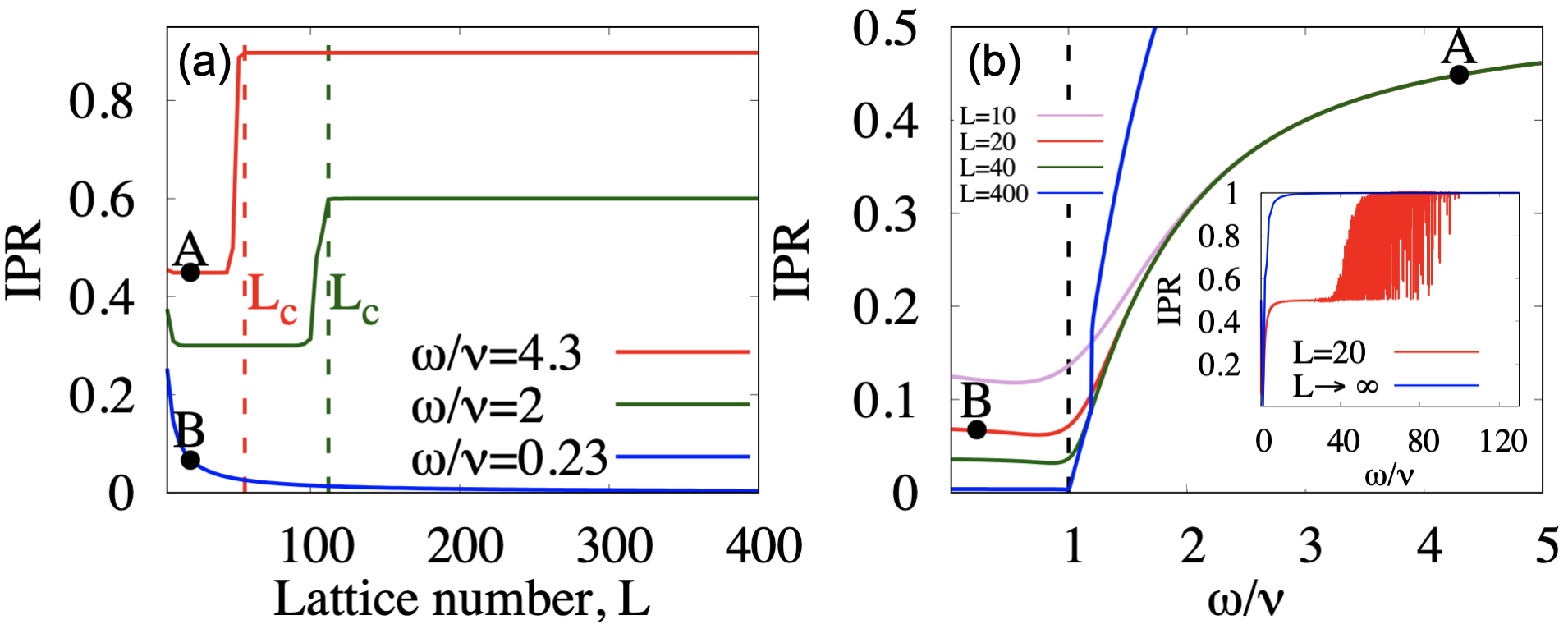}
\caption{(a) The Inverse participation ratio (IPR) on the lattice number $L$  for non-trivial topological states with $w/v = 2.0$ and $4.3$, in green and red colors, respectively. An abrupt change  can be identified  in the IPR index, marked by  the transport length $L_c$.
As a comparison, in (a) the curve of trivial case with $w/v = 0.23$ only gives an exponential scaling factor, i.e., the  curve in blue color.
For a fixed system size $L = 10$, $20$, $40$ or $400$, in (b) the  IPR index is depicted as a function of $w/v$, with the large $w/v$ limit shown in the inset.
The markers $\text{A}$ and $\text{B}$ correspond to our experimental conditions.
}
\end{figure}

\noindent {\it Wave localization.}\textemdash
In order to give a quantitative figure of merit in verifying localization-delocalization transition in these edge states, we calculate the corresponding inverse participation ratio (IPR) index by defining~\cite{IPR},
\begin{eqnarray}
\text{IPR} \equiv \frac{\sum_{i=1}^L |\Psi_i|^4}{(\sum_{i=1}^L |\Psi_i|^2)^2},
\end{eqnarray}
which represents a measure of the number of sites contributing to a given state. 
The clear distinction of IPR index indicates the characteristic features of SSH model, providing a criterion to distinguish the extended (delocalized) states from the localized ones. 

Moreover, the IPR index for non-trivial topological states reveals  an abrupt change at a characteristic length, denoted as $L_c$, as shown in Fig. 2(a).
When the system size is smaller than $L_c$, the IPR index for the edge modes is small (IPR $<0.5$), indicating the existences of  symmetric or anti-symmetric combinations of the edge modes. Specifically, in this regime we have 
\begin{eqnarray}
\text{IPR}=\frac{[1+(\frac{w}{v})^{-2}][1-(\frac{w}{v})^{-2N}]}{2\, [1-(\frac{w}{v})^{-2}][1+(\frac{w}{v})^{-2N}]}.
\end{eqnarray}
Only when $L > L_c$, the IPR index jumps up by a factor of 2, indicating that only strongly isolated states at one side are supported. 
As a comparison, for all the trivial cases we do not have such an abrupt change in the IPR curve, such as $w/v = 0.23$ illustrated in Fig. 2(a), which only gives an exponential scaling factor.

As shown  in Fig. 2(b), such an abrupt change in the IPR index also exists for a fixed system size when we vary the value of $w/v$.
For different lattice sizes, such as $L = 10$, $20$, or $40$,  the corresponding IPR index curves merge into a single curve; while IPR for $L=400$ is doubled when $w/v$ is sufficiently large. \\

\begin{figure}[t]
\includegraphics[width=8.4cm]{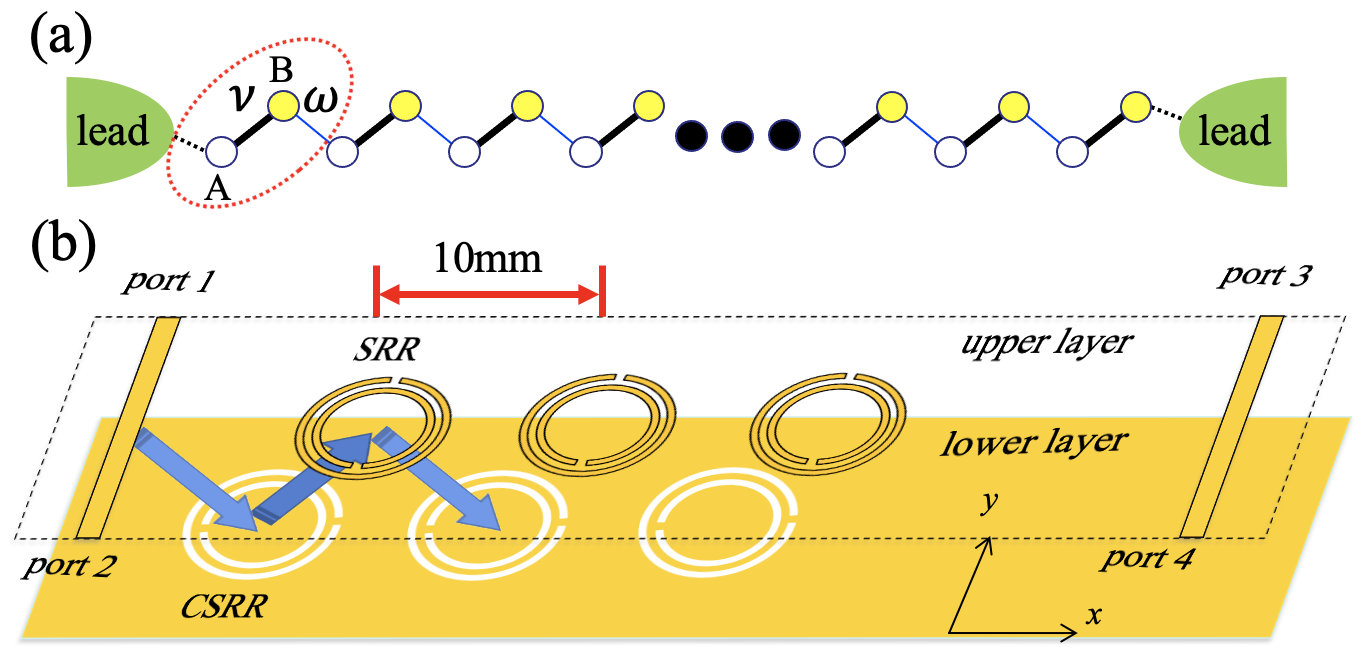}
\caption{
(a) Schematic of SSH model for 1D microwave propagation in dimer lattices, denoted as  $A$ and $B$. The red circle marks the unit cell of ``diatomic basis''. 
(b) The implementation of the SSH model is realized in a finite-size chain by combining split-ring resonator (SRR) and complementary split-ring resonator (CSRR).
By varying the orientation of SRR and CSRR, we can manipulate the coupling strengths for the  intra- and inter-cells, denoted as $v$ and $w$, respectively.
The two leads in the front and end nodes shown in (a) are implemented by two transmission lines in (b), denoted with the input port $1$,  reflection port $2$, and transmissions ports $3$ and $4$, respectively.
}
\end{figure}

\noindent {\it Experiments.}\textemdash
To demonstrate the transport between two edge states even with a lattice size much larger than the penetration depth, as the schematic shown in Fig. 3(a), we fabricate a 1D chain composited by a finite number of sites in the  diatomic basis, denoted as $A$ and $B$. 
The implementation of SSH model is realized by combining  SRRs and CSRRs together, but fabricated in the front and back layers, respectively,  as illustrated in Fig. 3(b).
By bonding these two interleaved layers close enough, with a controllable separation, an effective 1D array can realize the  SSH model.
In addition to the dimer unit, we manipulate the intra- and inter-cell coupling strengths by varying the orientation of SRRs and CSRRs, respectively.

Regarding the materials, our 1D SSH array is fabricated on the substrate Roger $4003C$, which has the  thickness $1.6$ mm and with copper used as the metallic part of $0.035$ mm thickness on it.
The resonators are designed with $7.6$ mm in diameter, $0.4$ mm in line width, along with $0.4$ mm between two rings,  $0.4$ mm for the ring gaps, and $10$ mm for the lattice constant.
For the $L = 20$ sample, we have $106.2$ mm in total length from the left lead to the right one.
It is known that one can manipulate the coupling strengths between SRRs by angle  rotations~\cite{Ser09, See17, Jiang}. 
Moreover, for the coupled SRR and CSRR, the texture of orientation produces a great impact on the inter-resonator coupling strength~\cite{Lin17}.
For a fixed angle of rotation in CSRR, the inter-resonator coupling strength can be varied from $30$ MHz to  $400$ MHz with respect to the angle of rotation in SRR.

Moreover, in order to perform direct measurements on the transmission and reflection spectra,   we also introduce  two leads in the front and end nodes, as illustrated in Fig. 3(a).
With the help of two transmission lines, as shown in Fig. 3(b),  here, we define the input (port $1$),  reflection (port $2$), and transmissions ports ($3$ and $4$).
To extract the magnitude of coupling strengths in our diatomic cell,  simulation results obtained by finite-element simulations are also applied to fit the experimental data, in order to  estimate the values of ratio between inter- and intra-cell coupling strengths, i.e., $w/v$.

With the help of experimental parameters extracted from our previous work~\cite{Lin17}, in the following we demonstrate the direct observation of non-trivial topologically protected edge states for microwave propagation in our 1D chain.
Here,  an even number of the total lattice sites is fabricated, i.e.,  $L = 20$, for two sets of orientations of SRR and  CSRR at the angles:  $(210\degree, 290\degree)$ for non-trivial case and  $(30\degree, 110\degree)$ for  trivial case, respectively.
As the selected angles in these two cases are supplementary, the coupling strengths $w$ and $v$ are simply interchanged, with the estimated values  of $w/v = 0.23$ and $4.3$, respectively. \\

\begin{figure}[t]
\includegraphics[width=8.4cm]{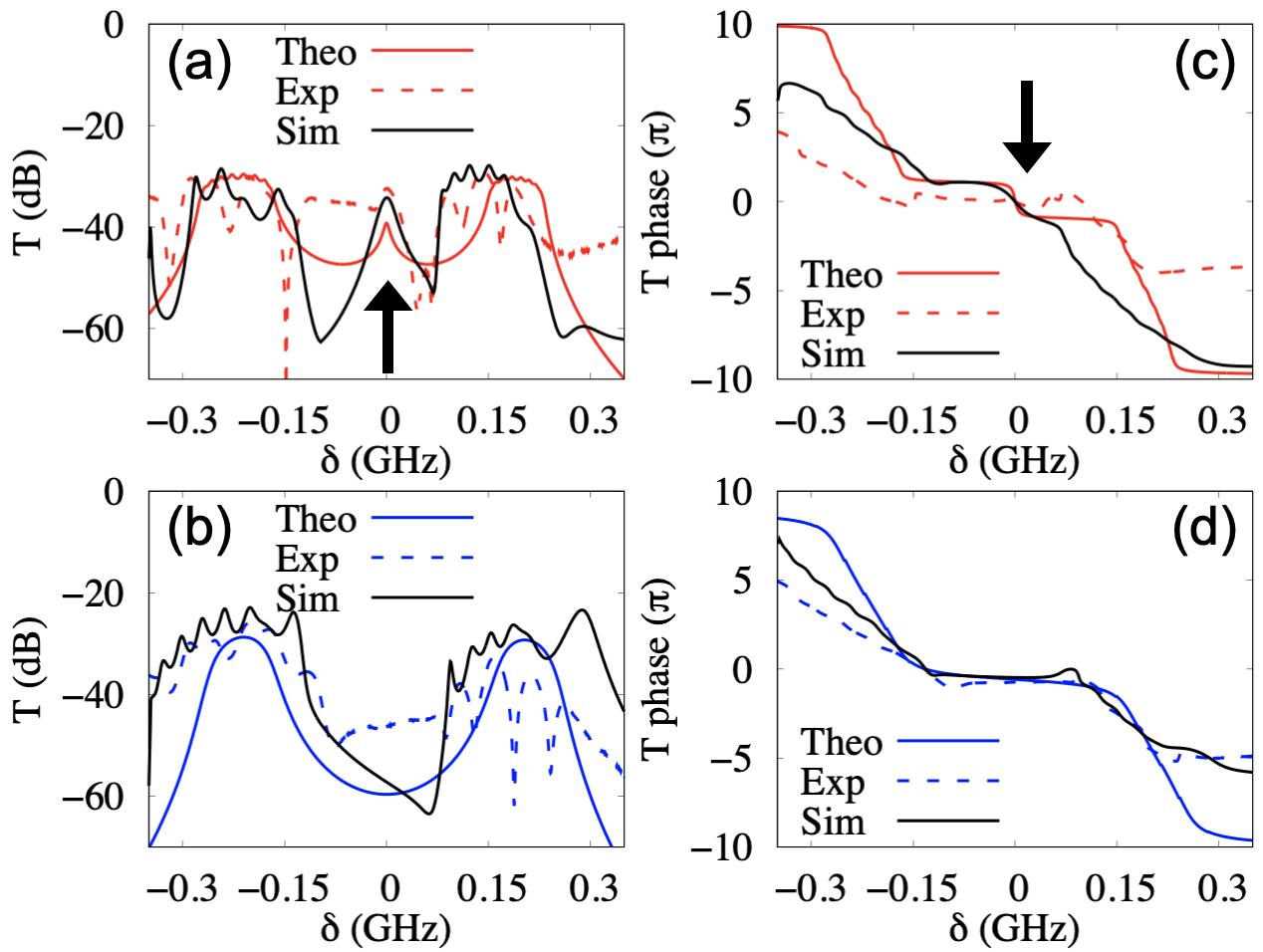}
\caption{(a) Transmission spectra $|T|$ (in dB scale) of 1D SSH model obtained from the theoretical model, experimental measurements and simulations in non-trivial ($w/v = 4.3$) case, while those in trivial case ($w/v = 0.23$) is shown in (b). (c) The corresponding transmission phase (in unit of $\pi$) obtained from the theoretical model, experimental measurements and simulations in non-trivial ($w/v = 4.3$) case; while those in trivial case ($w/v = 0.23$) is shown in (d). 
The arrows depicted in (a) and (c) indicate the peak and phase shift in the non-trivial case.
}
\end{figure}

\noindent {\it Transmission Spectroscopy.}\textemdash
In response to directly probe supported non-trivial topologically protected edge states experimentally through the transmission spectroscopy, in Fig. 4(a), we illustrate the calculated transmission spectra $|T|$ (in dB scale) of 1D SSH model obtained from the theoretical model with non-trivial ($w/v = 4.3$) case for $L=20$, together with the transmission spectrum $S_{31}$ obtained by direct measurements and finite-element simulations. Non-trivial topological edge states, i.e., the zero-modes, produce the additional resonance peak at gap center ($\delta=0$). Even though fabrication errors introduce inevitable discrepancy in the measured spectrum, a clear resonant tip can be seen in the transmission spectra, indicating the robustness of topologically protected edge states. On the other hand, transmission spectra in trivial ($w/v = 0.23$) case only presents two passband as in Fig. 4(b).

In addition to the transmission amplitudes, the corresponding phase changes are also shown in Figs. 4(c) and 4(d), for the non-trivial and trivial cases, respectively.
One can see that near the zero-detuning region, when $w/v  > 1$, a phase jump of $2 \pi$ can be clearly identified in Fig. 4(c), for the existence of the non-trivial topologically protected edge state from two Zak phases in $\pm \pi$ at $\delta=0$. As a comparison no such a phase jump happens  for the trivial topological phase, as shown in Fig. 4(d). These transmission spectra clearly demonstrate that our observed states indeed are non-trivial zero modes.

\begin{figure}
\includegraphics[width=8.4cm]{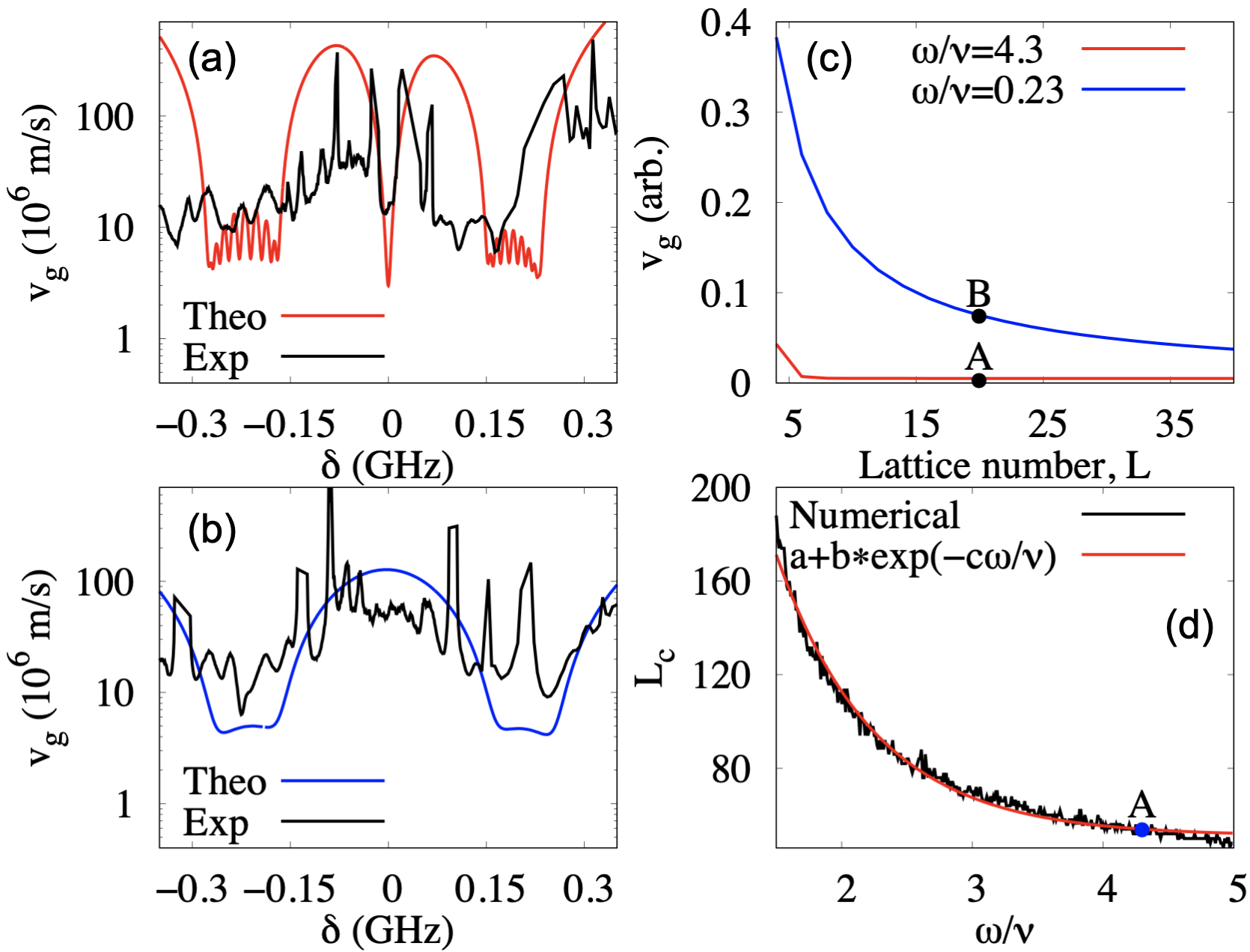}
\caption{The measured group velocities for (a) non-trivial case $w/v = 4.3$ and (b) trivial case $w/v = 0.23$, as a function of frequency detuning $\delta$.
Here, in addition to the theoretical prediction,  the experimental measurements are performed by pulse excitations.
(c) On the zero-detuning condition $\delta = 0$, we depict the group velocity as a function of the lattice number $L$.
One can see that the group velocity for non-trivially topological states $w/v = 4.3$, depicted in red-color, significantly remains as a constant when the system size is smaller than the characteristic length, i.e., $L < L_c$. However, the group velocity  for the non-trivially topological states $w/v = 0.23$, depicted in blue-color, decays exponentially to the system size. 
The  characteristic length $L_c$ as a function of $w/v$ is also depicted in (d). 
The marker $\text{A}$  corresponds to our experimental conditions.
} 
\end{figure}

With pulse excitations, depicted in solid-black curves in Figs. 5(a) and (b), we show the direct experimental measurements on the group velocity of supported edge states, for non-trivial case $w/v = 4.3$ and trivial case $w/v = 0.23$, respectively. 
In addition to the theoretical prediction,  a significant dip in the measured group velocity, as well as in the simulated one, can be clearly identified for the non-trivially case, indicating a much lower group velocity than the light speed.
Our results give a direct observation of the topologically protected edge states in slow-light, with the measured group velocity $\sim 13.6 \times 10^6$ m/s.
On the contrary, the group velocity of trivial cases propagates at the speed of light in the bulk, i.e., $\sim 0.75 \times 10^8$ m/s with the refractive index $4$.

In order to manifest the transport property of our supported non-trivial edge states in the finite-size SSH chains, we show the corresponding group velocity on 
 the zero-detuning condition $\delta = 0$ in Fig. 5(c), as a function of the lattice number $L$.
 As  one can see, the group velocity for the non-trivially topological states $w/v = 4.3$, depicted in red-color, significantly remains as a constant when the system size is smaller than the characteristic length, i.e., $L < L_c$.
As a comparison, the group velocity for the trivially topological states $w/v = 0.23$, depicted in blue-color, is exponentially decaying to the system size.

With the introduction of this characteristic length $L_c$, in Fig. 5(d) we provide a guideline on the maximum system size allowed to transport the edge states in the SSH model.
Unlike the penetration depth $\xi$, the characteristic length $L_c$ revealed here is much larger than $\xi$. 
Counter-intuitively, as long as the system size is small than $L_c$, the transport between two edge modes is allowed  even for a system with its size much larger than the penetration length $L \gg \xi$. 
For a better understanding its origin, $L_c$ is associated with a threshold in the coupling energy given by $\frac{\Delta}{v}(L_c)=10^{-16}$. 
From the expression of the energy gap $\Delta$, empirically, one finds that: 
\begin{eqnarray}
L_c=2\frac{\ln(10^{16}) + \ln[\frac{\frac{w}{v}}{(\frac{w}{v})^2-1}]}{\ln(\frac{w}{v})}.
\end{eqnarray}
In more practical situations, disorder or dissipation may attribute to the transport threshold such that  $\Delta(L_c)\sim E_{th}$, in which $E_{th}$ is disorder strength or relaxation rate~\cite{dissipative}.
\\

\noindent {\it Conclusion.}\textemdash
On the contrary to the common belief that  the wavefunctions of supported  edge states in a finite-size chain should decay exponentially in the bulk, 
we reveal the existence of a transport length  in the bulk size for edge states in  1D SSH chains, which is much larger than the penetration depth of the edge modes.
A non-exponential scaling factor to the system size is demonstrated though an abrupt change in the wave localization, indicating a possible phase transition for the topological edge states in a finite-size SSH chain.
Regarding experimental demonstrations, we implement a chain of  interleaved split-ring resonators  and complementary split ring resonators with a proper setting in the orientation texture.
The existence of non-trivially topological edge states, which are zero modes, is not only  demonstrated through the transmission spectra, but  also verified through the observation of a $2\pi$  phase jump.
With the non-zero coupling energy in a finite-size system,  the group velocity of  microwave propagation is directly measured through pulse excitations, revealing  a very slow group velocity, down to $\sim 0.01$ of light speed in free space. 
The transport length revealed and verified in our experiments provides a crucial guideline for  utilizing photonic topological devices, which demonstrates the possibility to transport optical information with topological protection~\cite{T1, T2, T3, T4, T5}.

\section*{Acknowledgement}
We are grateful to the National Center for High-performance Computing for computer time and facilities. Fruitful discussions with C. C. Huang and W. H. Chang are acknowledged. G.W. acknowledges the financial supports from National Natural Science Foundation of China (11604231); Natural Science Foundation of Jiangsu Province (BK20160303). 
This work is financially supported by the Ministry of Science and Technology, Taiwan under grant Nos.106-2112-M-005-007, 108-2923-M-007-001-MY3 and No. 109-2112-M-007-019-MY3), Office of Naval Research Global, US Army Research Office, and Research Center for Sustainable Energy and Nanotechnology, NCHU. This work was co-funded by European Union - PON Ricerca e Innovazione 2014-2020 FESR/FSC - Project ARS01\_00734 QUANCOM.

\end{document}